\documentclass[superscriptaddress,prl,twocolumn,amsmath,amssymb]{revtex4-1}
\usepackage{graphicx}% Include figure files
\usepackage{dcolumn}% Align table columns on decimal point
\usepackage{bm}
\usepackage{eurosym}
\usepackage{amsmath}
\usepackage{amssymb}
\usepackage{epstopdf}
\usepackage{dcolumn}% Align table columns on decimal point
\usepackage{bm}% bold math
\graphicspath{{../images/}}
\usepackage{soul}
\usepackage{braket}
\def\>{\rangle}
\def\<{\langle}

\def\togli#1{}
\def\labell#1{\label{#1}}

\def\section#1{{\par\em #1:--- }}
\usepackage[bottom]{footmisc}

\begin{document}
%\fbox{{\scriptsize Preliminary draft.\today}}

\title{Entanglement-assisted quantum metrology}

\author{Zixin Huang}

\affiliation{ School of Physics, University of Sydney, NSW 2006, Australia }

\author{Chiara Macchiavello}

\affiliation{Dip. Fisica and INFN Sez. Pavia, University of Pavia, via Bassi 6, I-27100 Pavia, Italy}

\author{Lorenzo Maccone}

\affiliation{Dip. Fisica and INFN Sez. Pavia, University of Pavia, via Bassi 6, I-27100 Pavia, Italy}

\begin{abstract}
  Entanglement-assisted quantum communication employs pre-shared
  entanglement between sender and receiver as a resource. We apply the
  same framework to quantum metrology, introducing shared entanglement
  between the preparation and the measurement stage, namely using some
  entangled ancillary system that does not interact with the system to
  be sampled. This is known to be useless in the noiseless case, but
  was recently shown to be useful in the presence of noise. Here we
  detail how and when it can be of use.
    For example, surprisingly it is useful when randomly time sharing two channels where ancillas do
   not help (depolarizing).
   We show that it is useful for all levels of
  noise for many noise models and propose a simple optical experiment to test
  these results.
\end{abstract}
%\pacs{}
%quantum mechanics, 03.65.Ta
%quantum algorithms and protocols, 03.67.Ac
%quantum optics, 42.50.Gy
%Quantum noise, 42.50.Lc

\maketitle

Entanglement-assisted communication 
\cite{sd,shor,shor1} employs pre-shared entanglement between sender
and receiver in addition to the signals sent through the channel.
This doubles the capacity of a noiseless channel, as illustrated in the 
well known superdense coding protocol \cite{sd}, and is even
more beneficial in the presence of noise \cite{shor1,dc-noise}.  
Here we apply the same framework to quantum metrology
\cite{qmetr,rev1,phase-comp,rev2,dowling2008quantum}, which studies how 
quantum effects may aid parameter estimation.  

\begin{figure}[h]
\includegraphics[trim = 1cm 1cm 1cm 2cm, clip, width=0.9\linewidth]{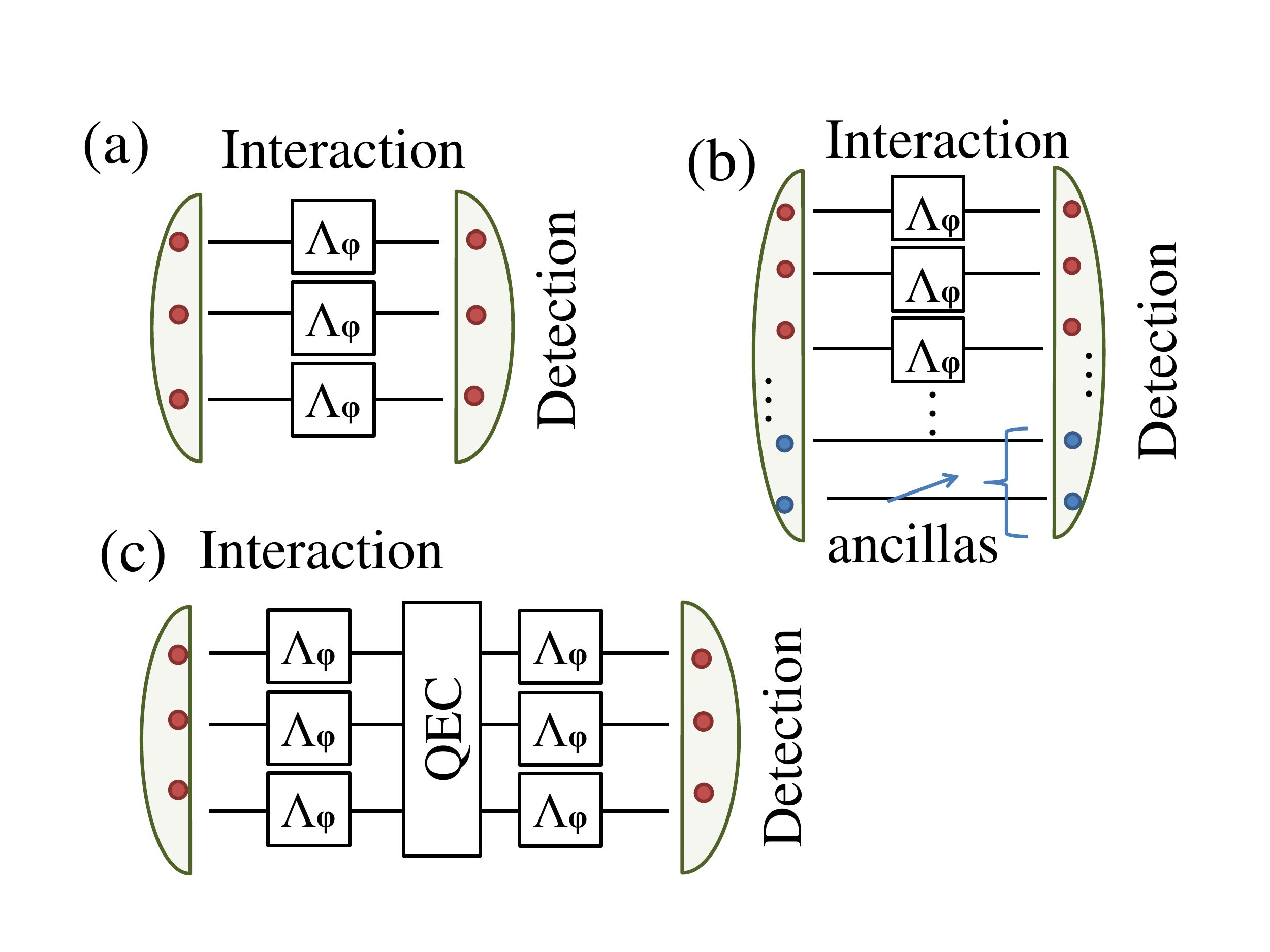}
\caption{\label{f:schema}(a) Conventional quantum parameter estimation: probes are
  prepared in a joint entangled state, they interact (independently)
  with the probed system through a noisy map $\Lambda_\varphi$, and
  they are jointly measured. (b) Entanglement-assisted parameter
  estimation: ancillary systems are employed that do not interact with
  the probed system. (c) QEC schemes: after the interaction
  the errors are corrected and, possibly, the probes interact again.
%  \comment{In Fig. (c) we should remove the ancilla states, as it was done 
% in Lorenzo's drawing} 
%
}
\end{figure}

A quantum parameter estimation is composed of three stages
(Fig.~\ref{f:schema}a): the preparation stage, where some probe
systems are initialized; the sampling stage, where the probes interact
with the system to be sampled (this interaction encodes the parameter
on the probes); the measurement stage, where the probes are measured
and the outcome is processed to yield the parameter estimate.
Entanglement-assisted quantum metrology (Fig.~\ref{f:schema}b) refers
to the scenario in which the probes are entangled with an ancilla that
does not participate to the sampling stage (similar 
ideas have also been studied in the context of noisy channel estimation \cite{fujiwara, fujiwara2}). 
Then at the measurement stage a joint measurement is performed between probes and ancilla. 
 It was recently realized that this is useful in the presence of noise
\cite{rafal,qeckraus,qecarrad,qeckessler,
  haine,kolodynski2013efficient}, although it was known to be useless
in the noiseless case \cite{qmetr}. Here we detail how entangled
ancillas can be used, what are the gains one can achieve and which
explicit measurement strategies can achieve such gains, analyzing the
most important qubit channels. We also propose a simple experiment
that can test our results.

The entanglement-assisted scenario should not be confused with the
generic use of entanglement in quantum estimation, that has been
extensively studied previously,
e.g.~\cite{caves,caves1,qmetr,nph,dinani,shaji,chiara,ian,chavez,cillis},
showing how entangled probes achieve better precision than unentangled
ones.  It is also different from the application of quantum error
correction to metrology \cite{qeckraus,qecarrad,qeckessler,qecozeri}
where all the systems are involved in the interaction
(Fig.~\ref{f:schema}c), although some scenarios analyzed in Refs.
\cite{qeckraus,qecarrad,qeckessler,qecozeri} go beyond this conventional
quantum error correction scheme.
Instead, in the entanglement-assisted
scenario, the ancilla does not interact and can be considered as
noiseless, if we suppose (as is often the case) that the noise is
relevant especially during the interaction stage. This translates into
a reduced noise acting on the global state and a reduced resource
count: indeed in quantum metrology the resource count refers to the
number of times that the probed system is sampled \cite{rev2}. So, the
entangled ancillas should be accounted for as separate resources, as
is done in quantum communication \cite{sd,shor,shor1}.

We will detail the increase in achievable precision in the presence of
an entangled ancilla with respect to the one achievable in its
absence, instead of dealing with the issue of whether one can beat the
standard quantum limit or achieve the Heisenberg bound. Indeed, it is
known that the Heisenberg scaling cannot be achieved asymptotically
for many of these noise models
\cite{guta,davidovich,jarzna,durkin,alipour}, although it can often be
achieved in the non-asymptotic regime \cite{brauns,alipour1,geo600}.
Interestingly, it has been pointed out
\cite{qeckraus,qecarrad,qeckessler} that the Heisenberg scaling {\em
  can} be recovered through entanglement-assisted metrology even
asymptotically in the very special case of orthogonal noise, even
though only a lower scaling is achieved in the absence of ancillas
\cite{chavez}.  However a general analysis of entanglement-assisted
metrology was lacking up to now: our results show the attainable
precision enhancement for the case of the Pauli channels (including
the fully depolarizing case) and of the amplitude damping.  Moreover,
dephasing and erasure noise do not allow for any enhancement in the
entanglement-assisted scenario, although entanglement among the probes
is helpful \cite{rafal}. Our analysis then concludes the study of
entanglement-assisted estimation for all relevant qubit channels.

Most previous literature studies the precision through the quantum
Cramer-Rao (QCR) bound, with the promise that such a lower bound to
precision is significant because it is asymptotically achievable.
However, since we want achieveability in the non-asymptotic regime, we
must provide estimation strategies and prove that they achieve the
QCR, at least in a feedback scenario \cite{feedb1,feedb2}. Indeed,
both the QCR and the error in the employed strategy may depend on the
unknown parameter to be estimated \cite{gdurkin}, but a feedback
strategy converges exponentially to the ``sweet spot''
\cite{dowling1}, so that it has only a logarithmic cost in terms of
resources \cite{feedb1}. Even though the rigorous way to determine
phase errors is the Holevo variance \cite{holevo,ph-gen}, as is usually done
in the literature we will use the customary variance, since the two
match for sufficiently precise estimation strategies.

The QCR \cite{holevo,helstrom,caves,caves1} is a lower bound to the
precision $\Delta\varphi$ of the estimation of a parameter $\varphi$:
under reasonable hypotheses, $\Delta\varphi\geqslant 1/\sqrt{\nu
  J(\rho_\varphi)}$ where $\nu$ is the number of times the estimation
is repeated, and $J$ is the quantum Fisher information (QFI)
associated to the global state $\rho_\varphi$ of probes and ancillas
(after the interaction $\Lambda_\varphi$ with the probed system). The
QFI is
\begin{eqnarray}
J(\rho_\varphi)=\sum_{j,k:\lambda_j+\lambda_k\neq 0}
{2|\<j|\rho'_\varphi|k\>|^2}/({\lambda_j+\lambda_k})
\labell{fishin}\;,
\end{eqnarray}
where $\rho'_\varphi=\partial\rho_\varphi/\partial\varphi$,
$\lambda_j$ and $|j\>$ are the eigenvalues and eigenvectors of
$\rho_\varphi$. The map $\Lambda_\varphi$ encodes the phase parameter
$\varphi$ onto the probes: $\rho_\varphi=\Lambda_\varphi[\rho]$, where
$\rho$ is the initial state. Without loss of generality, for qubit
channels we suppose that the phase is encoded onto the computational
basis by the unitary $U_\varphi=|0\>\<0|+e^{i\varphi}|1\>\<1|$. For
the sake of simplicity we will consider situations where the noise
maps ${\cal E}$ acts after $U_\varphi$, namely $\Lambda_\varphi={\cal
  E}\circ U_\varphi$. If the noise map and $U_\varphi$ do not commute,
this is an important restriction of our analysis (required to make the
problem tractable, as nontrivial effects arise otherwise
\cite{chavez}), but it is not a restriction for the erasure, amplitude
damping and depolarizing noise models which commute with $U_\varphi$.

To find the best bound, one must maximize the QFI, which in general
depends both on the input state $\rho$ and on the unknown parameter
$\varphi$. The former optimization depends on the noise map, the
latter can be taken care of using feedback mechanisms
\cite{feedb1,feedb2,gdurkin,dowling1}. For all noise maps, we can use
the convexity of QFI \cite{fujiwara} to choose a pure input state
$\rho=|\psi\>\<\psi|$.  Indeed, supposing that
$\rho=\sum_j\lambda_j|j\>\<j|$, we have
$J(\rho_\varphi)=J(\sum_j\lambda_j\Lambda_\varphi[|j\>\<j|])\leqslant
\sum_j\lambda_jJ(\Lambda_\varphi[|j\>\<j|])$. (The QFI is not convex
if $\lambda_j$ depended on $\varphi$, but an extended convexity still
holds \cite{alipour1}.)

\section{Amplitude damping}
\begin{figure}[hbt]
\includegraphics[trim = 0cm 0cm 0cm 0cm, clip , width=0.8\linewidth]{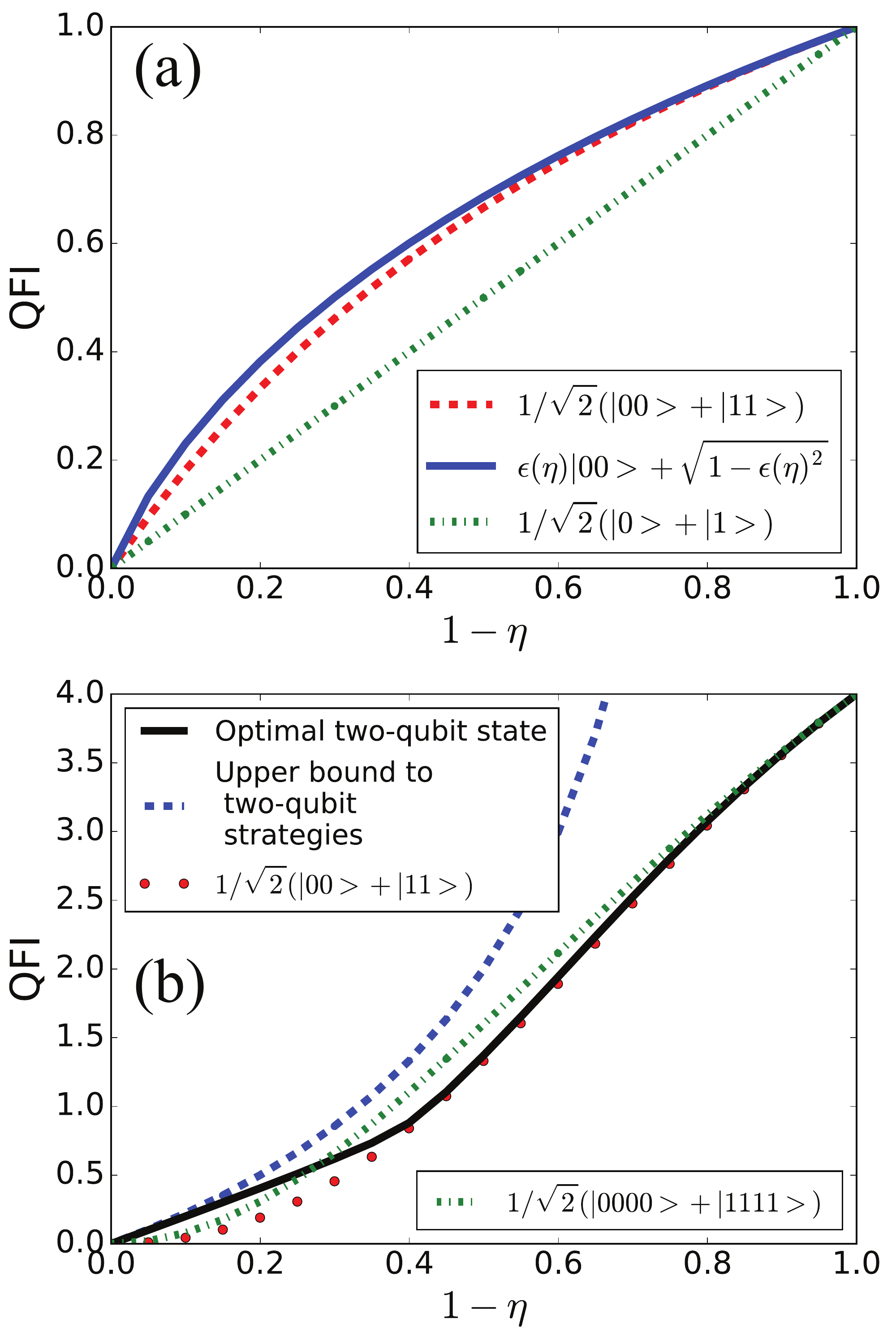}
\caption{\label{fig:adcsingle} Amplitude damping channel. (a) Single
  qubit probe: QFI of the optimal ancilla-assisted state
  $|\psi_\gamma\>$ (continuous), the maximally entangled state of
  probe and ancilla (dashed) and single qubit state (dotted). For all
  $\eta$ the ancilla-assisted strategies have better QFI. (b)
  Two-qubit probes: the optimal two-qubit state (continuous line) and
  the ``noon'' state (which is optimal only in the noiseless case, red
  circles) are compared to a genuinely entangled four-qubit state (two
  probe and two ancilla qubits, dot-dashed line). We also plot the
  upper bound to the ancilla-less estimation with two probes from
  \cite{durkin} (dashed), which is achievable in the high noise regime
  (where the dashed and continuous lines superimpose). The
  ancilla-assisted strategy beats any entangled-probe strategies for
  $\eta < 0.7 $.  The state can be further optimized.}
\end{figure}

We start by analyzing the amplitude damping
channel with Kraus operators
\begin{equation}
A_0 = \left(
\begin{array}{cc}
 1 & 0 \\
 0 & \sqrt{1-\eta } \\
\end{array}
\right) ,\
A_1 = \left(
\begin{array}{cc}
 0 & \sqrt{\eta } \\
 0 & 0 \\
\end{array}
\right),
\end{equation}
where $\eta$ is the probability of decay $|1\>\to|0\>$. This map is
agnostic on the direction of the $x$ and $y$ axis of the Bloch sphere:
a rotation around the $z$ axis leaves $A_0$ unchanged and adds an
inconsequential phase factor to $A_1$. Thus we can optimize the
single-probe input state among the family
$\epsilon|0\>+\sqrt{1-\epsilon^2}|1\>$. The optimal state has
$\epsilon=1/\sqrt{2}$, and its QFI is $1-\eta$. To show that
entanglement-assisted strategies perform better, we compare this QFI
with the one of an entangled state of probe and ancilla
$|\psi_\gamma\>=\gamma|00\>+\sqrt{1-\gamma^2}|11\>$. This state might
not be the optimal state for the entanglement-assisted strategy, but
it outperforms the previous one. Indeed its corresponding output state
$(\Lambda_\varphi\otimes\openone)[|\psi_\gamma\>\<\psi_\gamma|]$ 
(where the map acts
only on the probe qubit and not on the ancilla) has a QFI of ${4
  (1-\eta )}/{(\sqrt{1-\eta}+1)^2}$ (also shown by \cite{kolodynski2013efficient}), if one optimizes over $\gamma$ for
each $\eta$.  Even the simple choice $\gamma=1/\sqrt{2}$ is
advantageous for all $\eta$ as its QFI is ${2(1-\eta)}/({2-\eta})$. These QFIs
are compared in Fig.~\ref{fig:adcsingle}a.

One observable that achieves the QCR bound for the last QFI is ${O} =
\Pi_\psi + 2\ket{\Phi^+}\bra{\Phi^+}$, with $\ket{\Phi^+} =
\frac{1}{\sqrt{2}}(\ket{00} + \ket{11})$ and $\Pi_\psi =
(\ket{01}\bra{01} + \ket{10}\bra{10})$. Indeed, measuring the
observable $O$, the error on $\varphi$ is
\begin{equation}
  \Delta \varphi^2 = \frac{\Delta O^2}{(d \braket{O}/ d
      \varphi)^2} = \frac{1-\eta/2-
      (1-\eta)\cos^2{\varphi}}{(1-\eta) \sin^2{\varphi}}
\label{eq:ADCancilla} .
\end{equation}
This expression is optimized for $\varphi = \pi/2$, where the
Cramer-Rao bound of the optimal single-probe state is beaten
performing Bell measurements between probe and ancilla. As discussed
above, we can perform the optimization $\varphi\to\pi/2$ using a
feedback strategy \cite{feedb1,feedb2,gdurkin,dowling1} that uses an
additional phase factor \cite{feedb1} in the interferometer to drive
the global phase to this ``sweet spot''.

The ancilla-assisted advantage persists also when we use entangled
probes. The case without ancillas was analyzed in \cite{durkin} where
an upper bound $N(1-\eta)/\eta$ to the Fisher information was given,
which is achievable in the high-noise regime $N\eta/(1-\eta)\gg
1$.

In the case of two qubit probes, one can also
perform a numerical optimization of the QFI on generic two-qubit
states (presented in Fig.~\ref{fig:adcsingle}b, solid line). 
 Interestingly,
the performance of the optimal state is similar to the ``noon'' state
$|\Phi^+\>$, which is known to be optimal in the noiseless case.
However, all these strategies can be beaten by using ancillas (as was
noted also in \cite{rafal}, although no explicit example was presented
there). In Fig.~\ref{fig:adcsingle}b, dotted-dashed line, we show the
performance of a the four-qubit NOON state
$(|0000\>+|1111\>)/\sqrt{2}$ of two probes and two ancillas:
it beats the optimal state of two probes for noise
levels $\eta \approx 0.7$, where its QFI is $\frac{8 (\eta -1)^2 \left(2 (\eta -1)^2 \cos (8 \varphi )+(\eta -2) \eta  ((\eta -2) \eta +2)+2\right)}{((\eta -2) \eta +2)^3}$, which is optimal for $\varphi = \frac{2\pi {n}}{8}$, 
where $n$ is an integer.
One observable that achieves the QCR for the single-qubit probe is
$O = 2 \ket{N}\bra{N} + \Sigma$, where $\ket{N} = 1/\sqrt{2}(\ket{0000} -\ket{1111})$ and $\Sigma = \ket{0011}\bra{0011} + \ket{0111}\bra{0111} + \ket{1011}\bra{1011}$.  Note that this particular four-qubit state is not necessary optimal.

% \begin{figure}[hbt]
%   \includegraphics[width=0.6\linewidth]{depolarising.pdf}
%   \caption{\label{fig:depolarising} Depolarising channel. For all of p
%   the ancilla-assisted strategy (solid line) has better QFI than an
%   unentangled probe (dotted line).}
% \end{figure}

\section{General Pauli}
The generalized Pauli channel is described by
\begin{equation}
\Lambda[\rho]=  (1- {p_1 - p_2 -p_3}) \rho + {p_1} \sigma_x \rho
\sigma_x ^\dagger 
 + {p_2 }\sigma_y \rho \sigma_y ^\dagger +  {p_3 }\sigma_z
 \rho \sigma_z ^\dagger, 
\label{eq:pauli}
\end{equation}
with $\sigma_x$, $\sigma_y$, $\sigma_z$ Pauli matrices, and $p_1$,
$p_2$ and $p_3$ probabilities.  Two special Pauli channels are the
dephasing and the depolarizing noise.  Dephasing noise corresponds to 
$p_1=p_2=0$ in \eqref{eq:pauli} and it was proved that ancillas offer no 
advantage over the unentangled probe \cite{rafal}: for both cases the 
optimal QFI of a single-qubit probe is $(1 - 2 p_3) ^2$.
Depolarising noise corresponds to having $p_1 = p_2 = p_3 \equiv p/4$ in
\eqref{eq:pauli}: it describes an isotropic loss of coherence
%  \comment{I changed the parametrization here to make it more consistent with the general case. 
% ZH: could you please change it also in the figure and check that all the 
% formulas below are still correct?}
. As was also shown in \cite{kolodynski2013efficient}, in this case ancillas do help: indeed for a single-qubit probe, the
optimal state is $1/\sqrt{2}(\ket{0}+\ket{1})$ where the QFI is
$(1-p)^2$, whereas the QFI for a probe maximally entangled with an
ancilla is $2(1-p)^2/(2-p)$, which is always greater
% (Fig.\ref{fig:depolarising})
.

This result is very surprising since the depolarizing channel can be seen as 
a time-sharing (with probability $1-p$) of a noiseless channel and a channel 
where the state of the probe is replaced by a maximally mixed state 
$\openone/2$
(useless for estimation). For both of these channels the use of an
ancilla gives no advantage: the ancilla becomes important only when
they are randomly time-shared.

One observable that achieves the QCR for the single-qubit probe is
$O = \ket{+}\bra{+}, \ket{+} = 1/\sqrt{2}(\ket{0}+\ket{1})$, and for
the ancilla-assisted case, $O = \Pi_\psi + 2\ket{\Phi^+}\bra{\Phi^+}$
 (identical with the one for the ADC).

Even though dephasing and depolarizing channels commute with the
unitary $U_\varphi$, in general the Pauli channel does not: we will
consider the case where the noise acts only after $U_\varphi$. For a
single qubit probe in the generic initial state
$\epsilon|0\>+\sqrt{1-\epsilon^2}e^{i\alpha}|1\>$, the QFI is

\begin{eqnarray}
  J^{(na)}_\varphi&=& 4 \epsilon ^2 \left(\epsilon ^2-1\right) [ {p_1} (2 - 4 {p_3}) + 4 {p_3} - 4 {p_3} ({p_2} + {p_3})\nonumber   \\
                  & & -1 - 2 {p_1}^2 - 2 (-1 + {p_2}) {p_2}
                  \labell{jna}\; \\  \nonumber
                  & & +2 ({{p_1}}-{{p_2}}) \cos (2 (\alpha
                  +\varphi )) ({{p_1}}+{{p_2}}+2
                  {{p_3}}-1)] 
\end{eqnarray}
($na$ stands for ``no ancilla''), which is maximized for
$\epsilon=1/\sqrt{2}$ for all $\varphi$, and (as before) one can
optimize over $\varphi$ (or, equivalently, $\alpha$) using a feedback
strategy. To prove that the presence of an ancilla is beneficial, we
consider a maximally entangled state
$\frac{1}{\sqrt{2}}(\ket{00}+\ket{11})$ of probe and ancilla. In this
case the QFI is
\begin{equation}
J^{(a)}_\varphi=\frac{({p_1}-{p_2})^2}{{p_1}+{p_2}}+\frac{({p_1}+{p_2}+2{p_3}-1)^2}{1-{p_1}-{p_2}}.
\labell{ja}
\end{equation}
% \comment{again there's the mysterious 2 in front of $p_3$, although
%   here it might perhaps be justified from the form of the entangled
%   state...} 
   
We performed a numerical search over the parameter space
${p_1},{p_2},{p_3}$. For all possible values of ${p_1} +{p_2} +{p_3}
\leq 1$, the expression in Eq.\eqref{ja} is larger than that of
Eq.~\eqref{jna} for all $\alpha$.

\begin{figure}[hbt]
\includegraphics[trim = 4cm 4cm 4cm 4cm, clip , width=0.8\linewidth]{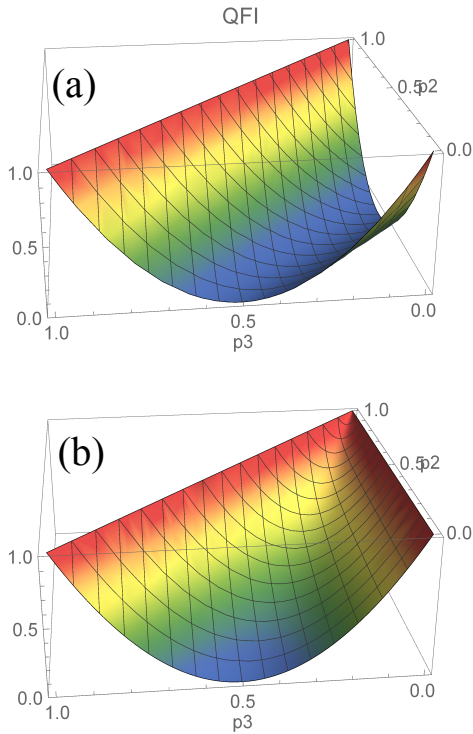}
  \caption{\label{fig:pauli} General Pauli channel (for the case
    $p_1=0$). (a) The optimized QFI for a single-probe in Eq\eqref{jna}, $\epsilon = 1/\sqrt{2},\alpha+\varphi = 0$  and (b) the QFI of the
    ancilla strategy in Eq. \eqref{ja}.     }
\end{figure}

   To illustrate this, the comparison between \eqref{jna} and \eqref{ja} is shown in
Fig.~\ref{fig:pauli}, for the case when  ${p_1} = 0$ and $\alpha+\varphi = 0$.
 The probe-ancilla observable $\hat{O} =
\ket{\Phi^-}\bra{\Phi^-} + \ket{\Psi^-}\bra{\Psi^-}$ achieves the QCR
bound relative to the QFI of \eqref{ja}, with $\ket{\Phi^-} =
\frac{1}{\sqrt{2}}(\ket{00} - \ket{11})$, $\ket{\Psi^-} =
\frac{1}{\sqrt{2}}(\ket{01} - \ket{10})$.

Note that when $p_1 ({p_2}) = 0$ and $p_1 (p_2) + p_3 = 1$, the QFI is
unity: this represents the case of orthogonal noise when the ancilla
strategy can recover the full information on the phase even in the
presence of noise, as was pointed out in
\cite{qeckraus,qecarrad,qeckessler}.

\section{Experiment proposal}
We propose here an experimental scheme to implement our proposal, based on the
use of a single photon, where two qubits are encoded in the path and 
polarisation degrees of freedom.

\begin{figure}[h]
\includegraphics[trim = 0cm 0cm 0cm 0cm , clip, width=1.0\linewidth]
{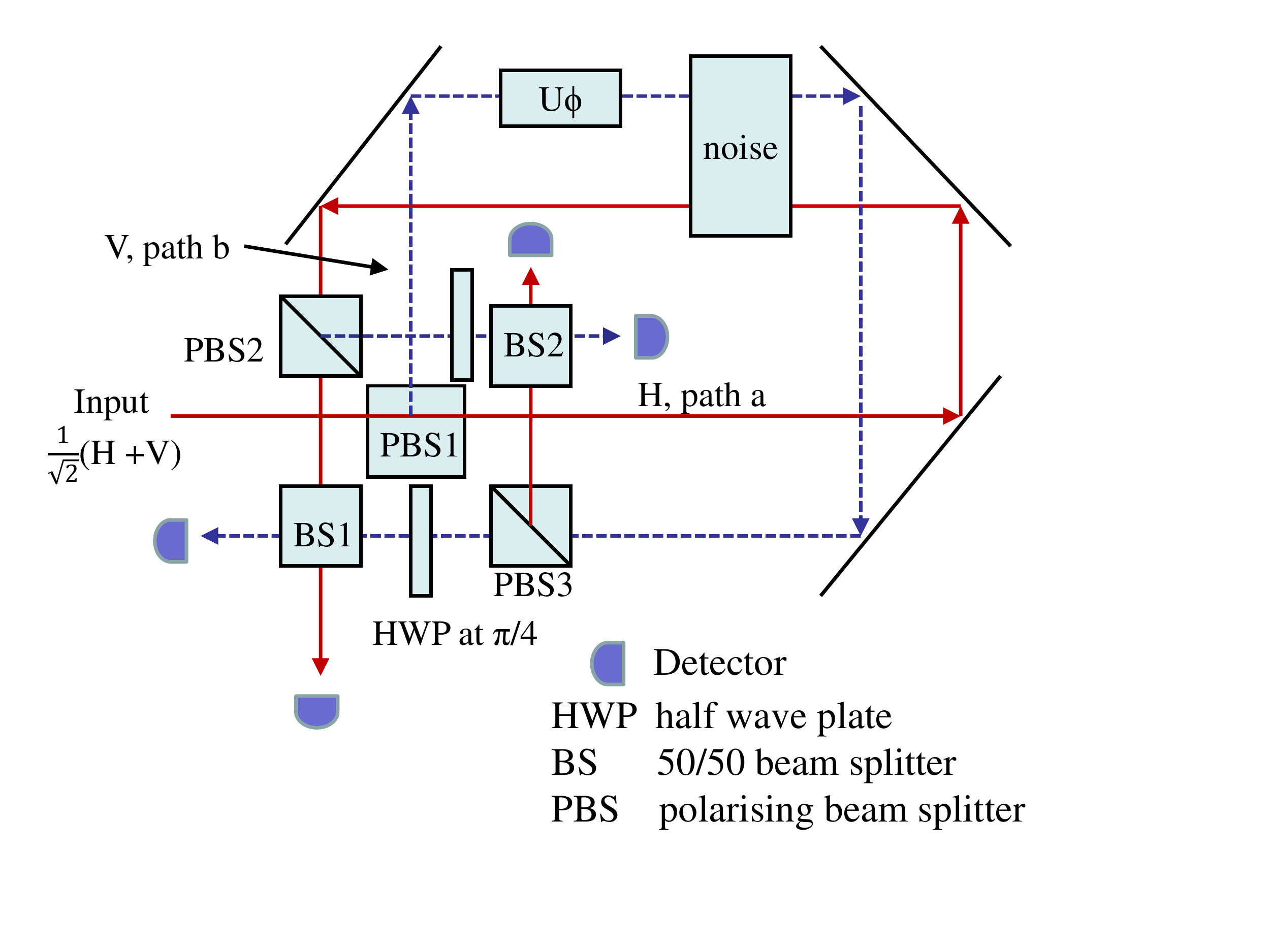}
  \caption{\label{fig:experimentalscheme} An experimental scheme to
    perform phase estimation using the ancilla-assisted scheme. See
    text for details.}
\end{figure}

Such a scheme, that can simulate the ancilla-assisted strategy, is
shown in Fig.\ref{fig:experimentalscheme}.  The experiment uses a
single photon which is prepared in a polarization-path entangled state
as follows and the noise acts on the polarization degree of freedom.
The initial polarisation state of the photon in this scheme is
$\frac{1}{\sqrt{2}}(\ket{H} + \ket{V})$. A polarizing beam splitter
(PBS) then transmits $\ket{H}$ and reflects $\ket{V}$. Therefore, PBS1
acts as an effective CNOT gate and puts into a polarization-path
entangled state $1/\sqrt{2}(\ket{H a} + \ket{Vb})$ (red solid = a,
blue dashed = b). The state is then transformed according to the
unitary $U_\varphi$ and the noisy channel. Different noise models can
be added, e.g.~using the techniques implemented in \cite{exps}. If the
bit is flipped, the flipped component is directed by PBS2 and PBS3
onto another path, which interfere at BS2.  The half wave plate (HWP)
rotates V polarization to H so that they interfere at the 50:50 beam
splitter (BS). The which-arm statistics after the BS are effective
projective Bell measurements in this basis.  That is, at BS1, the
outputs correspond to projecting onto $1/\sqrt{2}(\ket{Ha} \pm
\ket{Vb})$, and at BS2, $1/\sqrt{2}(\ket{Va} \pm \ket{Hb})$.  This
scheme is easily implementable with present-day technologies, e.g.~in
\cite{exps} similar schemes were experimentally realized and
controlled noise was introduced in an ancilla-assisted scenario for
different purposes.

\section{Conclusions}
In conclusion, we have studied the role of entangled ancillas in
metrology for the important classes of qubit noise models.  We have
shown that, for a single probe,
in the presence of amplitude damping, depolarizing noise as well
as general Pauli noise, 
an entanglement-assisted scheme provides an advantage in the efficiency
of phase
measurement over the unentangled case for all ranges of noise regimes.
We also derived the optimal
measurement procedures  which achieve the Cramer-Rao bound.

% \clearpage

We acknowledge useful feedback from Rafal Demkowicz-Dobrzanski.


\begin{references}
\bibitem{sd}C.H. Bennett, S.J. Wiesner, Phys. Rev. Lett. {\bf 69},
  2881 (1992).
\bibitem{shor} C. H. Bennett, P. W. Shor, J. A. Smolin, and
A. V. Thapliyal, Phys. Rev. Lett. {\bf 83}, 3081 (1999).
\bibitem{shor1} C. H. Bennett, P. W. Shor, J. A. Smolin, and A. V.
  Thapliyal, IEEE Trans. Inform. Theory {\bf 48}, 2637 (2002), Eprint
  quant-ph/0106052.
\bibitem{dc-noise}
Z. Shadman, H. Kampermann, C. Macchiavello and D. Bruss, 
New J. Phys.  {\bf 12}, 073042 (2010).
\bibitem{qmetr} V. Giovannetti, S. Lloyd, and L. Maccone, Phys. Rev.
  Lett. {\bf 96}, 010401 (2006).
\bibitem{rev1}V. Giovannetti, S. Lloyd, and L. Maccone, Science {\bf
    306}, 1330 (2004).
\bibitem{phase-comp} 
W. van Dam, G.~M.~D'Ariano, A. Ekert, C.~Macchiavello and M. Mosca,
Phys. Rev. Lett. {\bf 98}, 090501 (2007).
\bibitem{rev2}V. Giovannetti, S. Lloyd, and L. Maccone, Nature
  Photonics {\bf 5}, 222 (2011).
\bibitem{dowling2008quantum}J. P. Dowling, Contemporary physics {\bf
    49}, 125 (2008).

\bibitem{fujiwara} A. Fujiwara, Phys. Rev. A {\bf 63}, 042304 (2001); 
   A. Fujiwara, Phys. Rev. A {\bf 70} 012317 (2004).
\bibitem{fujiwara2} A. Fujiwara, H. Imai , J. Phys. A: Math. Gen. {\bf 36} 8093 (2003)


\bibitem{rafal} R. Demkowicz-Dobrzanski, L. Maccone, Phys. Rev.  Lett.
  {\bf 113}, 250801 (2014).

\bibitem{qeckraus}W. D\"ur, M. Skotiniotis, F. Fr\"owis, B. Kraus, Phys.
  v. Lett. {\bf 112}, 080801 (2014).
\bibitem{qecarrad} G. Arrad, Y. Vinkler, D. Aharonov, and A. Retzker,
  Phys. Rev. Lett. {\bf 112}, 150801 (2014).
\bibitem{qeckessler} E. M. Kessler, I. Lovchinsky, A. O. Sushkov, and M.
  D.  Lukin, Phys. Rev. Lett. {\bf 112}, 150802 (2014).
\bibitem{haine} S.~A. Haine,S. S. Szigeti, Phys. Rev. A {\bf 92}, 032317 (2015),
 S A. Haine, S S. Szigeti, M. D. Lang,C. M. Caves,  Phys. Rev. A {\bf 91} 041802, (2015)

\bibitem{kolodynski2013efficient} R. Demkowicz-Dobrza{\'n}ski, J. Kolodynski 
New Journal of Physics, \textbf{15}, 073043 (2013).



\bibitem{caves} S. L. Braunstein and C. M. Caves, Phys. Rev. Lett. {\bf 72},
3439 (1994).
\bibitem{caves1} S. L. Braunstein, C. M. Caves, and G. Milburn, Annals
  Phys. {\bf 247}, 135 (1996).
\bibitem{nph}G. Y. Xiang, B. L. Higgins, D. W. Berry, H. M. Wiseman,
  G. J. Pryde, Nature Photonics {\bf 5}, 43 (2011).
\bibitem{dinani} H. T. Dinani and D. W. Berry, Phys. Rev. A {\bf 90},
  023856 (2014).
\bibitem{shaji} A. Shaji and C. M. Caves, Phys. Rev. A {\bf 76}, 032111
(2007).
\bibitem{chiara} S. F. Huelga, C. Macchiavello, T. Pellizzari, A.K.
  Ekert, M.B. Plenio, J.I. Cirac, Phys. Rev. Lett. {\bf 79}, 3865
  (1997).
\bibitem{ian}M. Kacprowicz, R. Demkowicz-Dobrzanski, W. Wasilewski, K.
  Banaszek, I. A. Walmsley, Nature Photonics {\bf 4}, 357 (2010).
\bibitem{chavez} R. Chaves, J. B. Brask, M. Markiewicz, J.
  Kolody\'nski, A. Ac\'in, Phys. Rev. Lett. {\bf 111}, 120401 (2013).
\bibitem{cillis} L. Maccone and G. De Cillis, Phys. Rev. A {\bf 79},
  023812 (2009).
\bibitem{qecozeri}R. Ozeri arXiv:1310.3432 (2013).
\bibitem{guta}R. Demkowicz-Dobrzanski, J. Kolodynski, M. Guta, Nature
  communications {\bf 3}, 1063 (2012).
\bibitem{davidovich}B. Escher, R. de Matos Filho, and L. Davidovich,
  Nature Physics {\bf 7}, 406 (2011).
\bibitem{jarzna}M. Jarzyna, R. Demkowicz-Dobrzanski, New J.  Phys.
  {\bf 17}, 013010 (2015).
\bibitem{durkin}S.I. Knysh, E.H. Chen, G.A. Durkin, arXiv:1402.0495
  (2014).
\bibitem{alipour}S. Alipour, M. Mehboudi, A.T. Rezakhani Phys. Rev.
  Lett. {\bf 112}, 120405 (2014).
\bibitem{brauns}S. L. Braunstein Phys. Rev. Lett. {\bf 69}, 3598
  (1992).
\bibitem{alipour1}S. Alipour, A. T. Rezakhani Phys. Rev. A {\bf 91},
  042104 (2015).
\bibitem{geo600}R. Demkowicz-Dobrzanski, K. Banaszek, R. Schnabel
  Phys. Rev. A {\bf 88}, 041802(R) (2013).
\bibitem{feedb1}D. W. Berry, H. M. Wiseman Phys. Rev. Lett. {\bf 85},
  5098 (2000); D. W. Berry, H. M. Wiseman, J. K. Breslin Phys. Rev. A
  {\bf 63}, 053804 (2001); D. Berry, arXiv quant-ph/0202136 (2002).
\bibitem{feedb2}A. Hentschel, B.C. Sanders Phys. Rev.  Lett. {\bf
    104}, 063603 (2010); A. Hentschel, B.C. Sanders Phys. Rev. Lett.
  {\bf 107}, 233601 (2011).
\bibitem{gdurkin}G.A. Durkin, J.P. Dowling Phys. Rev. Lett. {\bf 99},
  070801 (2007).
\bibitem{dowling1}K.P. Seshadreesan, S. Kim, J. P. Dowling, H. Lee
  Phys. Rev. A {\bf 87}, 043833 (2013). 
\bibitem{holevo}A.S. Holevo, {\em Probabilistic and statistical
    aspects of quantum theory} (North Holland pub. co., Amsterdam,
  1982).
\bibitem{ph-gen} For the general case see also:
G.M. D'Ariano, C. Macchiavello and M.F. Sacchi, Phys. Lett. A {\bf 248}, 
103 (1998).
\bibitem{helstrom} C. W. Helstrom, ``Quantum Detection and Estimation
  Theory,'' (Academic Press, New York, 1976).


\bibitem{exps} A. Chiuri {\it et al}, Phys. Rev. Lett. {\bf 107}, 253602 
(2011); A. Orieux {\it et al},
Phys. Rev. Lett. {\bf 111}, 220501 (2013); A. Orieux {\it et al},
Phys. Rev. Lett.  {\bf 115}, 160503 (2015).



\end{references}
\end{document}